\documentclass[amsmat,amssymb,amsfonts,aps,prb,twocolumn,showpacs]{revtex4}

\usepackage{graphicx}
\usepackage{dcolumn}
\usepackage{bm}
\begin{document}

\title{Spin orbit   in curved  graphene ribbons}
 

\author{D. Gos\'albez-Mart\'inez (1) ,   J. J. Palacios(1,2), J. Fern\'andez-Rossier (1)}
\affiliation{(1) Departamento de F{\`i}sica Aplicada, Universidad de Alicante, San
Vicente del Raspeig, SPAIN \\
(2)  Departamento de F\'isica de la Materia Condensada, Universidad Aut\'onoma de Madrid}

\date{\today} 

\begin{abstract} 

We study the electronic properties of electrons in
flat and curved zigzag graphene  ribbons
using a  tight-binding model within the Slater Koster approximation.  
   We find that curvature dramatically enhances the action of spin orbit effects  in graphene ribbons and has a strong effect on the spin orientation of the edge states:  
  whereas  spins are normal to the surface in the case of flat ribbons, this is no longer the case in the case of curved ribbons. We find that for the edge states, the spin density lies always in the plane perpendicular to the ribbon axis, and deviate strongly from the normal to the ribbon, even for very small curvature and  the small spin orbit coupling of carbon. 
   We  find that curvature results also in an effective second neighbor hopping that modifies the electronic properties of zigzag graphene ribbons.  We discuss the implications of our finding in the spin Hall phase of curved graphene Ribbons. 

\end{abstract}

\maketitle
\section{Introduction}

The electronic structure of graphene  depends  both on  its structure at the atomic scale,  determined by the $sp^2$ hybridization, and on its structure at a much  larger
 length-scale,   determined the shape of the sample\cite{Dresselhausbook,Kane-MeleNT}. Thus,  the electronic properties of flat graphene differ in subtle but important ways from rippled graphene\cite{Meyer07}  and the properties of carbon nanotubes are fully determined by the way they fold\cite{Dresselhausbook}.  Curvature is believed to affect transport\cite{Mozorov06,Morpurgo06}, magnetic\cite{Brey08}  and spin relaxation properties of graphene\cite{Huertas06}.  
  In a wider context,  the interplay between mechanical deformations and electronic properties, the so called {\em flexoelectronics}, is giving rise to a new branch  in nanotechnology.  Whereas conventional electronics devices are  based on the capability to tune  their   working properties  by  application of external perturbations in the form of electric and magnetic fields,  mechanical deformation can have a major impact on  the properties of nanoelectronic devices. This results in a wide range of new effects, like piezoelectric nanogenerators \cite{Wang06} and field effect induced by piezoelectric effect\cite{ZnO-NL09} in ZnO nanowires  and  stress driven Mott transition in VO$_{2}$ nanowires \cite{Mott-NL09}. 

The spintronic and magnetic properties of graphene are intriguing.  From the theory side, there are two bold predictions. First, graphene should display a Quantum Spin Hall phase  with spin-filtered states in the zigzag edges\cite{Kane-Mele1,Kane-Mele2}.  Second, the same edges should have ferromagnetic order\cite{Fujita96,Son06,Cohen06,Gun07,JFR08,Yaz08,kim08,Fede09,Jung09,Soriano10}.  Whereas the  striking progress in the fabrication of atomically precise graphene ribbons has not permitted to confirm
 the presence of  stable zigzag edges yet\cite{Dai08,Cai10}, 
several groups have reported the observation of ferromagnetic order in graphene and graphite\cite{Esquinazi2002,Esquinazi2003,Wang09,Cervenka09,Esquinazi2010}, in most instances in samples that contain many flakes, structural disorder or have been irradiated.

The observation of the spin Hall phase in flat graphene  would require reducing the temperature below the spin-orbit induced gap, which is smaller than 10$\mu$eV \cite{Min06,Yao07,Gmitra09}.  From this point of view it would be desirable to  increase the strength of spin orbit interaction in graphene.  Hints of how this could be achieved come from experiments. On one side, the spin relaxation time of graphene, as measured in  lateral spin valves,  is in the range of 100 ps\cite{Tombros-Nature}, much shorter than expected from the small size of spin orbit  and hyperfine nuclear coupling \cite{YazyevNL}. Thus, some mechanism enhancing the strength of spin orbit interaction must be at play in these samples.  A possible candidate could be either curvature \cite{Huertas06,Kuemeth-Nature} induced by ripples and adatoms\cite{GraphaneGeim,Paco-Neto09}.  Curvature has been shown to enhance the effect of spin orbit coupling  in the case of carbon nanotubes, for which recent experimental  work has reported    zero field splittings induced by spin-orbit coupling splittings in the range of  200 $\mu$eV  \cite{Kuemeth-Nature}.

The  recently reported fabrication of curved graphene ribbons by unzipping carbon nanotubes\cite{Unzipped1,Unzipped2,Unzipped3} opens the  way towards the experimental study of the  effect of curvature on the edge states of graphene ribbons.  Here we study this system from the theoretical point of view and we compare the spin properties of a graphene ribbon both for flat and  curved  ribbons. 
In flat graphene the $\pi$ bands are decoupled from the $\sigma$ bands, unless spin orbit coupling is considered. 
However, the effect of spin orbit coupling on $\pi$ bands occurs only via virtual transition to higher energy $\sigma$ bands.

   In the case of flat graphene, it has been verified that the effect of spin orbit on the $\pi$ bands  can be properly described by an effective spin dependent second neighbor hopping  between  the $\pi$ orbitals. This is the so called Kane and Mele model\cite{Kane-Mele1,Kane-Mele2}, which  predicts that graphene is a Quantum Spin Hall insulator with a spin and valley dependent gap and peculiar 
spin-filter zigzag edge states\cite{Kane-Mele1,Kane-Mele2}. In the case of curved graphene, $\pi$ and $\sigma$ orbitals are coupled, and to the best of our knowledge the validity of the Kane-Mele model has not been tested.  This is why we adopt a different strategy\cite{Chico-Slater04,Min06,Huertas06,Saito09,Chico-Slater09} and we  use a four orbital tight-binding model, which includes both the $\pi$ orbitals as well as the $s$, $p_x$ and $p_y$ orbitals.

The rest of this manuscript is organized as follows. In section (\ref{method}) we describe the tight binding method used in our calculations and review some general results about the spin properties of the system .  In section (\ref{flat}) we present results for the electronic structure  of flat zigzag graphene ribbons and compare with those of the Kane-Mele  model.  In section (\ref{curved1}) we address the main point of this work, the electronic structure of edge states in  curved graphene zigzag ribbons.  In section (\ref{discussion}) we discuss our results.

\section{Formalism}
\label{method}
In this section we briefly comment the  two different tight-binding approximations used  to calculate the electronic structure and we provide some theory background. 

\subsection{Slater Koster approximation}

In most of the calculations in this work we use a multi-orbital approach, taking into account the 4 valence orbitals of the Carbon atom, $s,x,y,z$ similar to that used in previous work \cite{Chico-Slater04,Min06,Saito09,Chico-Slater09}. Thus, counting the spin, the single particle basis has 8 elements per carbon atom.  In addition, we passivate the edge carbon atoms with a single Hydrogen atom for which a single $s$ orbital, with the corresponding spin degeneracy, is included.  
The matrix elements of the Hamiltonian are computed according to the Slater Koster approach considering only first neighbor hoppings. For simplicity we approximate the overlap matrix as the unit matrix.
We model  both the carbon-carbon and carbon-hydrogen hoppings of graphene with a set of tight-binding parameters derived by Kaschner {\em et al.}\cite{Kaschner} from comparison with density functional calculations.  We show these parameters in table (\ref{SK-parameters}).

\begin{table}
\begin{center}
\begin{tabular}{ccccc}
   & $Vss\sigma$ & $Vsp\sigma$ & $Vpp\sigma$ & \ \ \ $Vpp\pi $ \\
\hline
\hline           
 C-C & -7.76 & 8.16 & 7.48 & -3.59\\
\hline
 C-H   & -6.84  & 7.81  &  & \\ 
\hline
\hline 
         & $\varepsilon_{s}^{C}=-8.8$& $\varepsilon_{p}^{C}=0.0$  &$ \varepsilon_{s}^{H}=-2.5$ & \\ 
\hline
\hline
\end{tabular}
\end{center}
\caption{On-site energies and Slater-Koster parameters involving the same atoms, carbon-carbon interaction and two different atoms, carbon-hydrogen interaction. All the values are in eV.}
\label{SK-parameters}
\end{table}

Spin orbit coupling is treated as an intra-atomic potential: 
\begin{equation}
{\cal V}_{SO}= \lambda \vec{S}\cdot\sum_I\vec{L}_I
\end{equation}
where $\lambda$ is the spin orbit coupling parameter, $\vec{S}$ is the spin operator and $\vec{L}_i$ is the orbital angular momentum operator acting upon the atomic orbitals of site $I$.  The representation of this operator in the basis $x,y,z$ is provided in the appendix(\ref{basis_set}). Whereas there is no consensus regarding the value of the atomic spin orbit coupling in carbon, the values reported in recent work range between $\lambda=4$ and $\lambda=8$ meV\cite{Yao07,Kuemeth-Nature,Min06}. In this work we  always discuss our results for values of $\lambda$ in that range and, when some physical insight is gained by so doing, for values of $\lambda$ much above the realistic range.

\subsection{One orbital tight-binding model}
The low energy physics of most graphene based nanostructures can be described with a tight binding model with a single orbital per atom, which can be taken as a $l=1$ atomic orbital projected along the local normal direction to graphene surface, the so called   $\pi$ orbitals. From the discussion above, it is apparent that the atomic spin orbit operator mixes orbital states in the same atom with different values of $m$.  However,  in some instances  is still possible to describe the low energy sector of graphene with  an {\em effective} Hamiltonian   governed by the by $\pi$ orbitals in which spin orbit gives rise to spin dependent hopping terms \cite{Kane-Mele1,Kane-Mele2}.
We  express the effective Hamiltonian using second quantization operators
 $c^{\dagger}_{I,\sigma}$ that create one electron in the atomic site $I$ with spin $\sigma$:
\begin{eqnarray}
&&{\cal H}_0=  \sum_{I,J,\sigma,\sigma'} {\cal T}_{I\sigma,J\sigma'} c^{\dagger}_{I,\sigma} c_{J,\sigma'} 
\label{Hamil1}
\end{eqnarray}
The Hamiltoniam matrix is the sum of four terms:
\begin{eqnarray}
{\cal T}_{I\sigma,J\sigma'}&=& t \delta_{\sigma,\sigma'} N^{(1)}_{I,J} +t'\delta_{\sigma,\sigma'} N^{(2)}_{I,J} +
\nonumber \\ &+&
 i t_{KM} \vec{\tau}_{\sigma,\sigma'}\cdot \left(\vec{d}_{1}\times\vec{d}_{2}\right) N^{(2)}_{I,J}
 \nonumber \\ &+&
i t_{R} \vec{n}\cdot\left(\vec{\tau}_{\sigma,\sigma'}\times \vec{d}_1\right) N^{(1)}_{I,J} 
\label{terms}
\end{eqnarray}
The elements of the matrix $N^{(1)}_{I,J}$  ( $N^{(2)}_{I,J}$ )  are equal to 1  when $I$ and $J$ are first (second) neighbor and zero everywhere else. Thus, the first two terms are the spin independent first and second neighobour hopping. The third term is  the Kane-Mele spin orbit model\cite{Kane-Mele1,Kane-Mele2}.
It is a spin dependent second neighbor hopping between sites $I$ and $J$ which have a common first neighbor $C$.
The unit vector along the bond between sites $I$ and $C$  ($C$ and $J$)  is denoted by $\vec{d}_{1}$ ($\vec{d}_{2}$).   In the Kane-Mele spin-orbit model the spin dynamics is linked to the bond orientation. Thus, in flat graphene and  graphene and graphene ribbons, the bonds lie in a plane so that the Kane-Mele spin orbit conserves the spin along the normal to the plane.  This is in contrast to the  curved ribbons and nanotubes considered below, for which the bond vectors are not restricted to a plane and no component of the spin operator is conserved.
The last term in equation (\ref{terms})  is the so called Rashba Hamiltonian, a first-neighbor spin dependent hopping.  In contrast to all the other terms, the Rashba explicitly breaks inversion symmetry so it is associated to an external electric field.  In the rest of this paper we calculate the band-structure of graphene based one dimensional structures using the
4 orbital Slater Koster model and compare with the results of the 
1 orbital model defined by eq. (\ref{Hamil1}) and (\ref{terms}), taking the Rashba term equal to zero. Whereas the 1 orbital model gives results very similar to those of the flat ribbon, this is not so in the case of curved ribbons.

 \begin{figure}
[hbt]
\includegraphics[width=0.9\linewidth,angle=0]{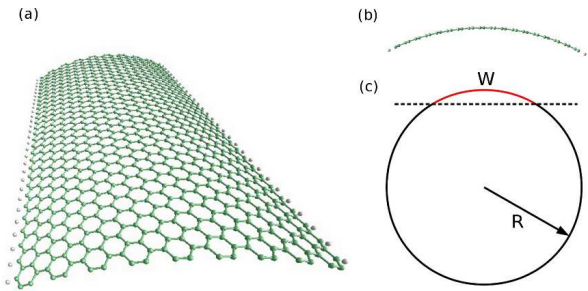}
\caption{ \label{fig1} Color online.  (a) Scheme of the procedure to generate curved ribbons. (b)  Section of a cruved ribbon.
(c) Perspective view of a curved graphene ribbon with edge atoms passivated with hydrogen.}
\end{figure}

\subsection{Some general results}
We study  one dimensional ribbons formed  repeating a crystal unit cell. Taking advantage of the crystal symmetry, the Hamiltonian can be written as  
$\sum {\cal H}_{n,m}(\vec k)$ where the indexes $n$ and $m$ run over the single particle spin-orbitals of the unit cell. For the one dimensional structures considered below,   the unit cell is shown in figure 1. The period of the crystal is given by the graphene lattice parameter, $a$,  $\vec{k}$ is given by a scalar $k$  and the Brillouin zone can be chosen in the interval $-\pi/a$ and $\pi/a$.  The eigenstates of the crystal are labelled with the band index $\nu$ and their crystal momentum $k$. They  are linear combination of the atom $I$  with quantum numbers $n=l,m$:
\begin{equation}
\Psi_{\nu k}(N,\vec{r}) = e^{i k N a} \sum_{I,n,s} C_{\nu k} (I,n,s) \phi_{n}(\vec{r}-\vec{r}_I) \chi_{s}
\end{equation}
where $N$ is an integer that labels the unit cell, and $ \phi_{n}(\vec{r}-\vec{r}_I)$ is the atomic orbital with orbital quantum numbers $n$ (which encodes  $l$ and $m$ ) in the atom $I$ of the unit cell.  The eigenstate of the spin operator  along the $z$ axis is denoted by $\chi_{s}$ where $s$ can take values $\pm \frac{1}{2}$. Because of the spin orbit coupling it is important to specify the positions of the atoms with respect to the spin quantization axis. 

Due to time reversal symmetry, every state with energy $\epsilon_{\nu}(k)$  must have the same energy that its time reversal partner, $\epsilon_{\nu'}(-k)$ where $\nu$ and $\nu'$ label states related by time reversal symmetry. In systems with  inversion symmetry the bands satisfy $\epsilon_{\nu}(k)=\epsilon_{\nu}(-k)$ so that, in the same $k$ point, there are at least two degnerate states.   In systems without inversion symmetry, like the curved ribbons considered below,  a twofold degeneracy at a given $k$ point is not warranted.  In this non-degenerate situation we can compute, without ambiguity,  the spin density associated to a given state with quantum numbers $\nu,k$ as:
\begin{equation}
\langle \vec{S}_{\nu,k}(I)\rangle  \equiv  \sum_{i,n,s,s'} 
C_{k,\nu} ^*(I,n,s) C_{k,\nu}(I,n,s')  \vec{S}_{s,s'}
\label{spindensity}
\end{equation}
where $\vec{S}_{\sigma,\sigma'}$ are the Pauli spin $1/2$  matrices.  

In the cases with inversion symmetry, like the flat ribbon  considered below,  for a given $k$ point there are at least  two degenerate bands.  Thus, any linear combination of states of the degenerate pair, $\Psi_{\nu k}$ and $\Psi_{\nu'k}$ is also eigenstate of the Hamiltonian and has different spin density.  In these instances, we include a infinitesimally small magnetic field in the calculation which breaks the degeneracy and permits to attribute a given spin density to a given state. When the calculated spin densities so obtained are   independent on the orientation of the infinitesimally small magnetic field, they  can be considered  intrinsic properties of the spin states. As we discuss below, this is the case of the spin filter states in flat spin ribbons, which point perpendicular to the ribbon and have a strong correlation between spin orientation, edge and velocity, as predicted by Kane and Mele\cite{Kane-Mele1,Kane-Mele2}.

In order to characterize the properties of a given state it will also be convenient to calculate their sublattice polarization:
\begin{equation}
\langle \sigma^z_{\nu,k}\rangle= \sum_{I,n,s} |C_{k,\nu} (I,n,s)|^2  \sigma_z(I) 
\end{equation}
where $\sigma_z(I)=+1$ when $I$ is a $A$ site and 
$\sigma_z(I)=-1$ when $I$ is a $B$ site.

\section{Flat graphene  zigzag ribbons}
\label{flat}

\subsection{Two dimensional graphene}
The spin properties and electronic structure of the flat ribbons considered below can be
related to those of the two dimensional graphene crystal. 
We briefly recall  the  the spin-orbit physics of two dimensional graphene as described within the Slater Koster model\cite{Min06}.   Within  this approach,  the electronic structure of  two dimensional graphene  is described by a 16 by 16 matrix, corresponding to the two atoms $A$ and $B$ of the unit cell \cite{Min06}.  
At zero spin orbit the 16 bands are two copies (one per spin), of 3 bonding-antibonding pairs of $\sigma$ bands and one bonding-antibonding pair of the $\pi$ orbitals which compose the states of the bands at the Fermi energy and are decoupled from the $\sigma$ bands. The Fermi surface is composed of two   points, $K$ and $K'$,  where the gap between the two $\pi$ bands vanishes.  In the neighborhood of both  $K$ and $K'$ points the two $\pi$ bands are linear and the $kp$ theory is formally identical to that of masless two dimensional Dirac electrons.    Spin orbit couples the $\sigma$ and $\pi$ orbitals, producing anti-crossings away from the Fermi energy and opening a gap at the $K$ and $K'$ points.  

Within the one orbital approach the Hamiltonian of graphene reads (see appendix (\ref{2D}) for details):
\begin{eqnarray}
{\cal H}= F(\vec{k}) \sigma_x  {\bf 1} + G(\vec{k}) s_z \sigma_z 
\end{eqnarray} 
where the $\sigma$ operators act on the sublattice space, $A$ and $B$,    ${\bf 1}$ is the unit matrix in the spin space and $s_z=\pm 1$ labels the spin. 
Here 
$F(\vec{k})$  is proportional to $t$ and 
 corresponds to first neighbor hopping  whereas  $G(\vec{k})$ is proportional to $t_{KM}$ corresponds to the Kane-Mele second neighbor spin orbit coupling. In the $K$ and $K'$ points the $F$ function vanishes but the spin orbit term gives rise to a gap $\Delta_{SO}= G(K)$.  As we show in the appendix, the function $g$ satisfies  $-G(K)=G(K')=  3\sqrt{3} t_{KM} $ .
Thus, because of the spin orbit interaction the  states at the Dirac points in graphene are described by the effective Hamiltonian  \cite{Kane-Mele1}:
\begin{equation}
h= \Delta_{SO} \tau_z \sigma_z s_z
\label{KM-Dirac}
\end{equation}
where $\tau_z=\pm1$ labels the valley quantum number,  $\Delta_{SO}= 3\sqrt{3}t_{KM}$.  Thus, for a given spin orientation $s_z=\pm1 $, spin orbit opens  a gap  that takes a valley dependent value $2 \tau_z s_z \Delta_{SO}=\pm2 \Delta_{SO}$.  This spin and valley dependent gap 
would  make   graphene  a peculiar type of insulator which  could not be connected with a standard insulator by smooth variation of a parameter in the Hamiltonian \cite{Kane-Mele1,Kane-Mele2}, i.e.,   a Quantum Spin Hall Insulator.

\subsection{Flat ribbons}
We now discuss the electronic structure of flat graphene ribbons with zigzag edges. The unit cell that defines the zigzag ribbon 
has $N$ carbon atoms and 2 hydrogen atoms that passivate the two carbon atoms in the edge of the ribbon. These structures were proposed by Nakada {\em et al} \cite{Nakada96}. Using the 1-orbital tight binding model, without SO coupling, they found that zigzag ribbons have almost flat bands at the Fermi energy, localized at the edges.  An important feature of zigzag edges  is the fact  that all the atoms belong to the same sublattice.  
Since the honeycomb lattice is bipartite,    a semiinfinite graphene plane with a zigzag termination must have
 zero energy edge states\cite{Palacios-JFR-Brey} whose wavefunction  decays exponentially in the bulk, with full sublattice polarization.
  In finite width zigzag ribbons,  the exponential tails of the states of the two edges hybridize, resulting in a  bonding-antibonding pair of weakly dispersing  bands\cite{JFR08}.
 The bands of the zigzag ribbon can be obtained by folding from those of two dimensional graphene, either with real or imaginary transverse wave vector, and  the longitudinal wave vector varying along the line that joins the two valleys, $K$ and $K'$. Thus, the valley number is preserved in zigzag ribbons.

  Spin orbit coupling, described with the 1 orbital model,  has a dramatic effect on  the (four) edge bands\cite{Kane-Mele1,Kane-Mele2} .  The second-neighbor hopping makes the single edge band dispersive, and overcomes the weak inter-edge hybridization. Interestingly the quantum numbers connect well with those at the Dirac points, that   are described by the effective Hamiltonian  (\ref{KM-Dirac}).   Thus, as we move from valley $K$ to $K'$  (positive velocity bands) the spin $\uparrow$  $(\downarrow)$ states of the edge with sublattice $A$  $(B)$ must change from the top of the valence band to the bottom of the conduction band.   The roles of spin and sublattice are reversed when considering the two bands that start at $K'$ and end at $K$. Thus  spin $\uparrow$ electrons move with positive velocity in one edge and negative velocity in the other.
\begin{figure}
[hbt]
\includegraphics[width=0.9\linewidth,angle=0]{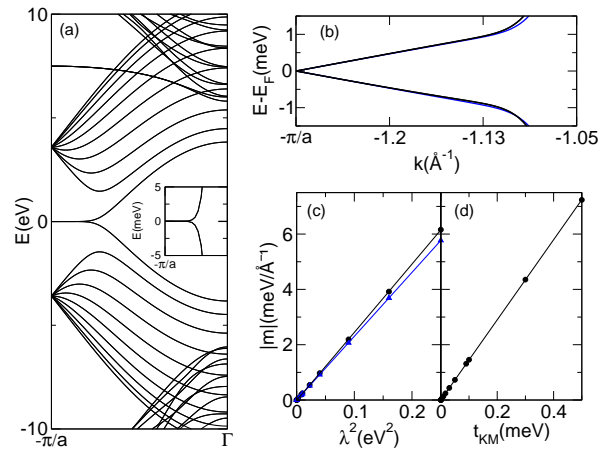}
\caption{ \label{fig2}  (a) Bands for flat ribbon without SO. Inset: Zoom of the edge bands  for case (a).
(b) Edge bands with SO ($\lambda=500$meV) and $t_{KM}=0.42$meV.  (c) Slope vs $\lambda^2$ (see equation ).  (d) Slope vs $t_{KM}$ }
\end{figure}

This scenario is confirmed by the 4 orbital model and is expected based upon the fact that 2D graphene with spin orbit coupling is a Quantum Spin Hall insulator.  In figure 2 we show the bands of a flat ribbon with $N=20$ carbon atoms. 
  In figure 2a we show the bands calculated without spin orbit coupling. The calculation shows both the edge and confined $\pi$ bands as well as some $\sigma$ bands higher in energy.  In the inset we zoom on the edge states to show that they are almost dispersionless  except when $k$ gets close to the Dirac point.  In figure 2b we show the same edge states, calculated with SO coupling, both within the 4 orbital and the 1 orbital model.  For this particular case we have taken $\lambda =500$ meV and $t_{KM}=0.42$ meV. Figure (\ref{fig2}c) shows a slightly different slope for valence and conduction bands for  the four orbital case. This electron-hole symmetry breaking  can not be captured with the one orbital Kane-Mele model. 
  
   It is apparent that the edge states acquire a linear dispersion $\epsilon =m k$. 
   The  slope $m$ of the edge states dispersion increases linearly with the size of the gap at the Dirac points, which in turn scales quadratically with $\lambda$ (and linearly with $t_{KM}$).  We can use $m$ to  quantify the effect of spin-orbit coupling on the edge states. 
Figure (\ref{fig2})d shows that, for flat ribbons, we can fit $m=\alpha \lambda^2$ with $\alpha\simeq 24 \frac{\AA}{ eV^2}$.

\begin{figure}
[hbt]
\includegraphics[width=0.9\linewidth,angle=0]{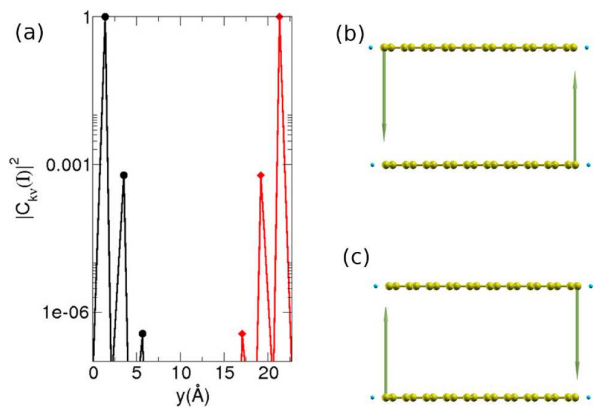}
\caption{ \label{fig3}  Properties of the edge state valence band. 
(a) Square of the Wave function for valence states with $k=+0.99\frac{\pi}{a}$. The sublattice polarization is apparent.
(b) Spin density for states with  $k=+0.99\frac{\pi}{a}$.
 (c)  Spin density of time reversal symmetric state with $k=-0.99\frac{\pi}{a}$.
The spin density of the 4 states is polarized perpendicular to  the sample.  
}
\end{figure}

 Since the flat ribbons have inversion symmetry, the bands have a twofold degeneracy. In order to avoid  numerical spin mixing of the degenerate states we apply a tiny  magnetic field (always less than 2 $\cdot$ 10$^{-4}$T) to split the states. As long as the associated Zeeman splitting is negligible compared to the spin orbit coupling, the direction of the field is irrelevant. By so doing, we can plot the spin density of the edge states without ambiguity.
 In figure (\ref{fig3}) we show both 
 the spin density of the four edge states with $k=0.99\frac{\pi}{a}$ and the square of the wave function for valence states, calculated with the 4-orbital model.
In agreement with the 1-orbital Kane-Mele model, the spin densities  are peaked in the edge, oriented perpendicular to the plane of the ribbon. We have repeated the calculation rotating the plane of the ribbon and obtained the same result. 
From inspection of figures (\ref{fig2}b) and 
(\ref{fig3}), it is apparent that the valence bands correspond to the edge states and display the spin filter effect, i.e., in a given edge right goers and left goers have opposite spin.

\section{Curved graphene zigzag ribbons}
\label{curved1}
The  calculation of the previous section , using the 4 orbital model, backs up the conclusions of the Kane-Mele 1 orbital model for the spin filter effect in graphene .  However, the bandwidth of the  edge states is less than 0.1 $ \mu eV$ for the accepted values\cite{Saito09} of $\lambda=5$meV. Thus,  the effect is very hard to observe in flat  graphene ribbons.     This leads us to consider ways to enhance the effect of spin orbit. For that matter, we calculate the edge states in a curved graphene zigzag ribbon, similar to those reported recently \cite{Unzipped1, Unzipped2,Chico09}, using the 4-orbital model.   In contrast to flat ribbons, the properties of the edge states of a curved ribbon can not be inferred from those of a parent two dimensional compound, because it is not possible to define a two dimensional crystal with a finite unit cell and constant curvature. 

The unit cell of the curved ribbons is obtained as fraction of a $(n,n)$ nanotube, with radius $R$ (see figure 1).  For a given nanotube we can obtain a series of curved ribbons with the same curvature $R^{-1}$ and different widths, $W$, or different number of carbon atoms $N$. 
We can also study ribbons with the same $N$  and different curvatures $R$, using a parent nanotube with different $n$.  Our curved ribbons are thus defined by $W$ and $R$ or, more precisely, by $n$ and $N$. 

\subsection{Energy bands}
\begin{figure}
[hbt]
\includegraphics[width=0.9\linewidth,angle=0]{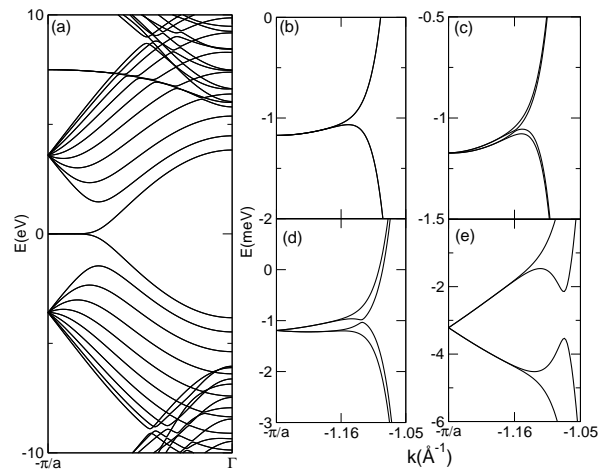}
\caption{ \label{fig4}  Electronic structure of curved zigzag ribbon with $R=4.1$nm. 
  (a) Bands for curved ribbon without SO. (b-e) Zoom of the edge bands  for $\lambda=0$ (b), 
  $\lambda =$5 meV (c) , $\lambda =$50 meV (d) and $\lambda =$500meV (e).}
\end{figure}

The energy bands  of curved ribbons is shown in figure (\ref{fig4}) for a ribbon with $N=20$ and $R=4.1$nm.  There are three  main differences with the flat ribbon. First, the edge states are dispersive even with $\lambda_{SO}=0$, as seen in figure (\ref{fig4})b. This effect can  be reproduced, within the one orbital model, including an effective second-neighbor hopping ($t'\simeq-1.5 $meV).  This dispersion  breaks electron hole symmetry and competes with the one induced by SO coupling, as seen in panels (\ref{fig4}c,d,e), with  $\lambda_{SO}=$5, 50 and 500 meV respectively.

\begin{figure}
[hbt]
\includegraphics[width=0.9\linewidth,angle=0]{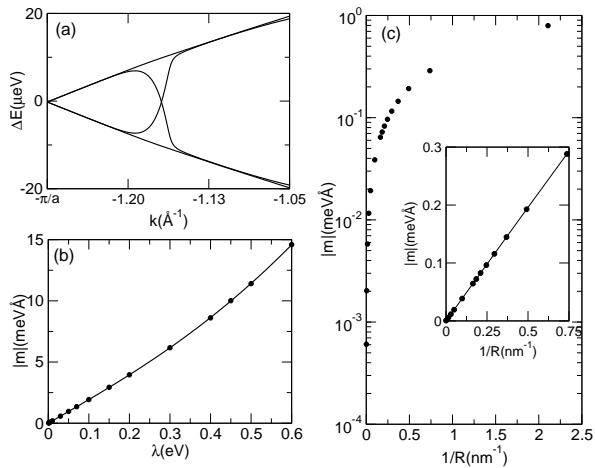}
\caption{ \label{fig6}
(a)  {\em Differential} energy bands  (see eq. \ref{relative} ) for  ribbons with $N=20$,  $R=4.1nm$ and $\lambda=5meV$
It is apparent that, for small $k$ the relative energy bands are linear, with slope $m$.  
(b) Slope of the relative energy bands as a function of $\lambda$, for $R=4.1$nm. (c) Slope of the {\em differential}  energy bands as a function of $\kappa=R^{-1}$, both in linear and logarithmic scale, for $\lambda=5$meV. A 100-fold  enhancement of the slope occurs in a very narrow range of small curvatures.   }
\end{figure}

Second, curvature enhances the effect of spin orbit coupling, as expected.  In order to separate the effect of spin orbit from the  effect induced by curvature, we define the {\em differential} bands
as the energy bands of the curved ribbon at finite $\lambda_{SO}$ subtracting the bands without spin-orbit:
 \begin{equation}
 \tilde{\epsilon}_{\nu}(k)\equiv\epsilon_{\nu}(k)-\epsilon_{\nu}(k,\lambda_{SO}=0)
 \label{relative}
 \end{equation}
  In figure (\ref{fig6})a we plot the {\em differential} bands for the ribbon with $N=20$ and $R=$4.1 nm.  They are 
two doubly degenerate linear bands with opposite velocities in the Brillouin zone boundary.  Thus, when the effect of curvature alone is substracted, the dispersion edge states look pretty much like those the flat ribbons
in the region close to the Brillouin zone boundary.  
Thus, we can also characterize them by the slope of the linear bands, $m$.  In figure (\ref{fig6})b we plot that  the slope of the edge bands as a function of $\lambda$ , obtained with the procedure just described,  for 
  the same ribbon discussed before ( $N=20$ ,  $R=$4.1nm).
 It is apparent that the slope $m$ is no longer a quadratic function of $\lambda^2$,  in contrast to the case of flat ribbons. Even more interesting, in figure (\ref{fig6}c) we plot the slope $m$ for a fixed value of $\lambda=$5 meV, as a function of the curvature $\kappa=R^{-1}$. We find a dramatic 100-fold increase at  small $\kappa$.
 This result is consistent with the effective $kp$ Hamiltonian  for carbon nanotubes \cite{Huertas06,Ando-curved}.

The curvature induced   enhancement of the spin-orbit effect  on the edge states of the curved ribbon  can be understood as follows. In flat graphene the Dirac bands are linear combination of atomic  $\pi$ orbitals with  quantum numbers $l=1$, $m=0$.  The effect of atomic spin orbit coupling, $\lambda \vec{L}\cdot\vec{S}$ can be understood perturbatively. To first order, spin orbit coupling has no effect on the product states $|\sigma\rangle\times |l=1,m=0\rangle$ states. Second order coupling, via intermediate states with $\Delta$ with respect to the Dirac point,  and orbital quantum numbers $l=1$, $m=\pm1$ results in an effective spin orbit Hamiltonian acting upon the $\pi$ orbitals, with strength $\frac{\lambda^2}{\Delta}$,   that conserves $S_z$.  Curvature changes this situation, because it mixes the $\pi$ orbitals with the $l=1,m\neq 0$ orbitals, resulting in an spin orbit Hamiltonian for the electrons at the Fermi energy \cite{Ando-curved,Huertas06} linear  in  the spin orbit coupling $\lambda$ . 

\begin{figure}
[hbt]
\includegraphics[width=0.9\linewidth,angle=0]{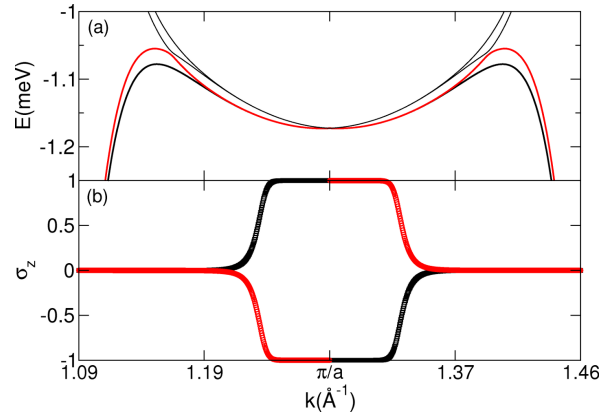}
\caption{ \label{fig5}
(Color Online) Detail of the edge states for curved ribbon with $R=4.1$nm  and $\lambda=5$meV.  
(a) Dispersion of the edge states (b) Sub-lattice polarization $\sigma_z$ as a function of  $k$ for the two lowest energy bands. }
\end{figure}

The third difference with the flat ribbon is apparent for the bands away from the zone boundary: they are not degenerate.
This is shown in figure (\ref{fig5})a, which is a zoom of figure (\ref{fig4})b, including states at both sides of the Brillouin zone boundary.   The lack of degeneracy is originated by the lack of the inversion symmetry of the curved ribbon.   Interestingly,  the degree of sublattice polarization, $\langle \sigma_z\rangle$,  shown in figure(\ref{fig5})b, anticorrelates with the splitting.   In other words, the states strongly localized at the edges are  insensitive to the lack of the inversion of the structure, which is non-local property.   

\subsection{Electronic properties}

\begin{figure}
[hbt]
\includegraphics[width=0.9\linewidth,angle=0]{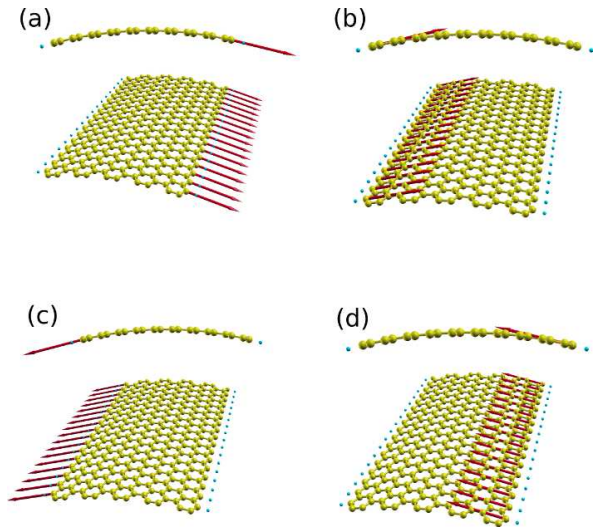}
\caption{ \label{fig7} 
Spin densities for states with $k=-\frac{\pi}{a}+0.01$ (panels (a) and (b)) and $k=\frac{\pi}{a}-0.01$ (panels (c) and (d)). }
\end{figure}

We now discuss the spin  properties of the edge states in the curved ribbons.  
 Notice that, due to the lack of inversion symmetry, there is no degeneracy at a given $k$ so that the spin density is an intrinsic property of the state.  
 In figure  (\ref{fig7})   we plot the magnetization density $\langle \vec{S}_{\nu,k}(I)\rangle$ for the two  lowest energy edge bands with momentum  $k=\frac{\pi}{a} + 0.01$  (upper panels) and $k=-\frac{\pi}{a} - 0.01$  (lower panel), for a ribbon with $R=4.1nm$ and $\lambda=5 meV$.   Whereas the correlation between the velocity, the spin orientation and the edge is the same than in flat ribbons, it is apparent that the quantization axis is no longer parallel to the local normal direction.  The spin of the edge states lies almost perpendicular to the normal direction. Thus, this is different from the case of nanotubes, where the spin quantization axis is parallel to the tube main axis, and different to the flat ribbon. 
 
\begin{figure}
[hbt]
\includegraphics[width=0.9\linewidth,angle=0]{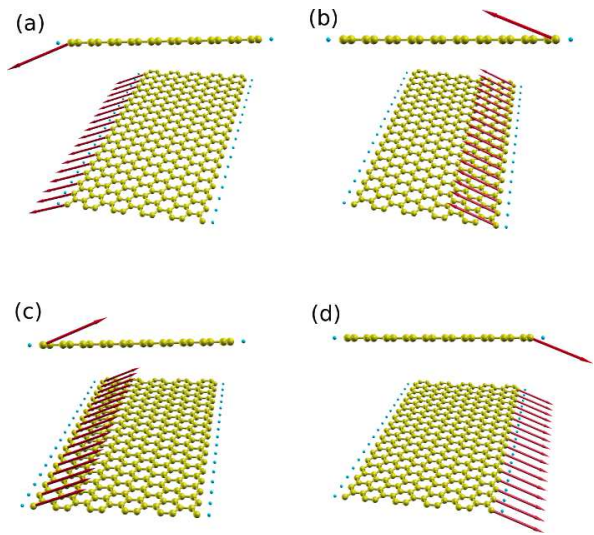}
\caption{ \label{fig8} 
Spin densities for states with $k=-\frac{\pi}{a}+0.01$ (panels (a) and (b)) and $k=\frac{\pi}{a}-0.01$  (panels (c) and (d)) in almost flat ribbon,  $R^{-1}=3.7\cdot10^{-3} nm^{-1}$.   Notice that curvature can not be appreciated.  }
\end{figure}

\begin{figure}
[hbt]
\includegraphics[width=0.9\linewidth,angle=0]{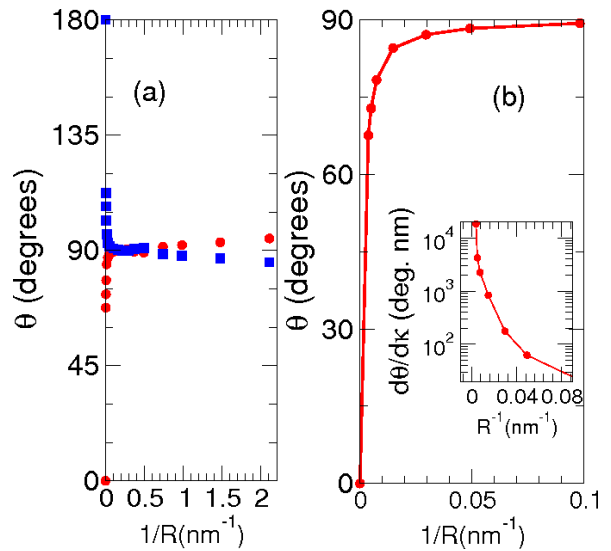}
\caption{ \label{fig9} 
Angle $\theta$ formed by the spin density and the local normal as a function of  curvature $\kappa=R^{-1}$  for
 $\lambda=5$ meV. Notice the extremely large $\frac{d\theta}{d\kappa}$.}
\end{figure}

 The effect is even more striking in the case of an almost flat ribbon, shown in figure (\ref{fig8}) for which the spin quantization direction is clearly not perpendicular to the ribbon plane.  Thus, a very small curvature  is enough to change the spin quantization direction of the edge states. This is better seen in (\ref{fig9}), where we plot the angle formed between the spin quantization axis and the local normal in the edge atom.  In (\ref{fig9})a we show the evolution of the angle for the states with $k=-\frac{\pi}{a}+0.01$ for the two lowest energy edge bands.  
   For flat ribbons $R^{-1}=0$, the angles are 0 and 180, {\em i.e.},  the quantization axis is pependicular to the ribbon plane.  In the opposite limit, for large curvature, the spin quantization angle lies perpendicular to the normal, i.e., tangential, but always in the plane perpendicular to the ribbon transport direction.

   The transition between the two limits is far from smooth. Even for the less curved ribbon that we have considered, with $R= 271 nm$, we have $\theta\simeq 70^0$. This is better seen in (\ref{fig9}b) where we show the low curvature region only for one of the bands.  The dramatic effect of curvature on the spin orientation of the edge states is quantified in the inset of figure (\ref{fig9})b where we show $\frac{d\theta}{d\kappa}$ as a function of curvature $\kappa$.  For small curvatures the  derivative blows up exponentially. 
     Thus, the spin orientation of the edge states is very sensitive to  moderate buckling deformation of the  ribbon.  This sensitivity is specially important for the small values of $\lambda$ adequate for carbon.  Larger values of $\lambda$  reduce the effect. 
   We have verified that the single orbital model with the generalized Kane-Mele Hamiltonian is not sufficient to capture the effect.

\section{Discussion and Conclusions}
\label{discussion}
We are now in position to discuss wether or not the spin filter effect could be more easily observed in curved graphene ribbons. One one side,  the bandwidth of the edge states is dramatically increased. For  moderate curvatures ($R=4.1 nm$) the bandwidth  of the edge states is  20$\mu$eV, and it can reach 60 $\mu eV$ for  $R\simeq 1nm$, to be compared with 0.1 $\mu eV$ for flat ribbons.  Thus, experiments done at 100 $mK$ could resolve the edge band of curved ribbons.    On the other hand, the curvature induced second neighbor hopping breaks  the electron-hole symmetry (see figure (\ref{fig5}a)), so that now there are 4 edge bands at the Fermi energy. At each edge there would be left goers and right goers with the same spin orientation, although slightly different $k$.  Taken at face value, this would imply that there is no spin current in the ground state and backscattering would not be protected. On the other hand, the lack of eh symmetry implies that the edges are charged. The  inclusion of electron-electron repulsion, even at the most elemental degree of approximation,  will probably modify the bands and restore the electron-hole symmetry. This is the case in most DFT calculations of zigzag ribbons.  Such a calculation is out of the scope of this manuscript.

In conclusion, we have studied the interplay of spin orbit coupling and curvature in the edge states of graphene zigzag ribbons. Our main conclusions are:
\begin{enumerate}
\item In the case of flat graphene ribbon, the microscopic 4-orbital model yields results identical to those of the Kane-Mele 1 orbital model. In particular, the edge states have the spin filter property \cite{Kane-Mele1,Kane-Mele2} and the spin is quantized perpendicular to the sample

\item Curved graphene ribbons also have spin-filted edge states states. The bandwidth of the edge bands of curved ribbons is increased by as much as 100 for moderate curvatures and is proportional to the curvature, for fixed spin orbit coupling. 

\item  Curvature induces a second neighbor hopping which modifies the dispersion of the edge states and, in this sense, competes with their spin-orbit induced dispersion.  

\item The spin of the edge states not  quantized along the direction normal to the ribbon. For moderate curvature, their quantization direction is a function that depends very strongly on the curvature of the ribbon. Above a certain curvature, the quantization direction is independent on curvature and perpendicular to both the normal and the ribbon direction.  The strong sensitivity of the spin orientation of edge states on the curvature suggest that flexural phonons can be a very efficient mechanism for spin relaxation in graphene. 

\end{enumerate}

{\it Note Added}: During the final stages of this work a related preprint has been posted\cite{lopezsancho2010}

\begin{center}{\bf ACKNOWLEDGMENT }\end{center}

We acknowledge fruitful discussions with David Soriano, A. S. N\'u\~ nez and S. Fratini.   This work has been financially supported by MEC-Spain (MAT07-67845,  FIS2010-21883-C02-01,  and  CONSOLIDER CSD2007-00010) and Generalitat Valenciana (ACOMP/2010/070).

\appendix

\section{Atomic orbital basis and matrix elements}
\label{basis_set}
In this appendix we give the expressions for the atomic orbitals and the corresponding angular momentum  matrix elements, necessary to compute the spin orbit matrix.  In our formalism,  the atomic orbitals are described in terms of the cartesian basis,  $s,p_{x},p_{y},p_{z}$, which is related to the basis of eigenstates of 
 $L^2$ and $L_z$,  through the following transformation: \cite{cohen-tanuji}
\begin{eqnarray}
| s \rangle&=&| l=0,m=0 \rangle\\
| p_{x} \rangle&=&-\frac{1}{\sqrt{2}} (| l=1,m=1 \rangle-| l=1,m=-1 \rangle\\
| p_{y} \rangle&=&\frac{i}{\sqrt{2}} (| l=1,m=1 \rangle+| l=1,m=-1 \rangle\\
| p_{z} \rangle&=&| l=1,m=0 \rangle
\label{solution}
\end{eqnarray}

The spin orbit Hamiltonian operator reads:
\begin{equation}
{\cal V}_{SO}=\lambda \left[ \frac{\hat{L}_{+}\hat{s}_{-}+\hat{L}_{-}\hat{s}_{+}}{2}+\hat{L}_{z}\hat{s}_{z} \right]
\label{H_so}
\end{equation}
which only affects the $l=1$ subspace.  
The matrix elements of this operator in the cartesian basis read:
\begin{eqnarray}
\begin{array}{ccccccc}
 & | p_{x}, \uparrow \rangle & | p_{y}, \uparrow \rangle & | p_{z}, \uparrow \rangle & | p_{x}, \downarrow \rangle & | p_{y}, \downarrow \rangle & | p_{z}, \downarrow \rangle \\ 
 | p_{x}, \uparrow \rangle  & 0 &-i\lambda/2 & 0 & 0 & 0 & \lambda/2 \\ 
 | p_{y}, \uparrow \rangle & i\lambda/2 & 0 & 0 & 0 & 0 & -i\lambda/2 \\ 
 | p_{z}, \uparrow \rangle & 0 & 0 & 0 & -\lambda/2 & i\lambda/2 & 0 \\ 
 | p_{x}, \downarrow \rangle & 0 & 0 & -\lambda/2 & 0 & i\lambda/2 & 0 \\ 
 | p_{y}, \downarrow \rangle & 0 & 0 & -i\lambda/2 & -i\lambda/2 & 0 & 0 \\ 
 | p_{z}, \downarrow \rangle & \lambda/2 & -i\lambda/2 & 0 & 0 & 0 & 0 \\ 
\end{array}
\end{eqnarray}

\section{Graphene in the 1 orbtal model}
\label{2D}
In this appendix we describe two dimensional graphene within the 1-orbital spin orbit model.  The two atom unit cell is taken dimer forming   degrees with the horizontal. The atoms are labelled $A$ and $B$. We use the  crystal vectors 
$\vec{a}_1=a\left(1, 0)\right) $ 
and $\vec{a}_2=a\left(Cos\frac{\pi}{3},Sin\frac{\pi}{3}\right)$.
%
The first neighbor hopping between the $R$ atom of a unit cell and its first neighbours reads
\begin{equation}
F(\vec{k})=t\left(1+e^{i \vec{k}\cdot\vec{a}_1}+e^{i \vec{k}\cdot\vec{a}_2}\right)
\end{equation}
which accounts for the fact that 1 of the firt neighbours is in the same cell. Two Dirac points at which $F(K)=F(K')=0$ vanishes are given by $Ka = (\frac{4\pi}{3},0)$ and $K'a=-Ka$. 

For the  Kane-Mele second neighbor spin-dependent hopping matrix elements involves coupling to 6 atoms, four of them in first neighbor cell and 2 of them in second neighbor cells. We have:
\begin{equation}
G(\vec{k})=it_{\rm KM}\left(
e^{i\vec{k}\cdot\vec{a}_1} -e^{i\vec{k}\cdot\vec{a}_2} +  e^{i\vec{k}\cdot(\vec{a}_2-\vec{a}_1)} -{\rm h. c. }
\right)\nonumber
\end{equation}
It can be readily verified that  $G(K'a)=-G(Ka)=3 \sqrt{3} t_{KM}  $.


\end{document}